\documentclass{millour}
\usepackage{wasysym}
\usepackage{graphicx}
\usepackage{hyperref}
\usepackage{natbib}

\def\Lsun{L$_{\rm \astrosun}$\,}

\TitreGlobal{What can the highest angular resolution bring to stellar astrophysics?}
\begin{document}

\title{Olivier Chesneau's work on massive stars} 
\runningtitle{Remembering O.~Chesneau} 
\newcommand{\folder}{.}


%
\author{F. Millour}\address{Laboratoire Lagrange, UMR7293,
  Universit\'e C\^ote d'Azur, Observatoire de la C\^ote d'Azur, CNRS,
  Bd. de l'Observatoire, 06300 Nice, France\\\email{fmillour@oca.eu}}

\begin{abstract}
Olivier Chesneau challenged several fields of observational stellar
astrophysics with bright ideas and an impressive amount of work to
make them real in the span of his career, from his first paper on
P~Cygni in 2000, up to his last one on V838~Mon in 2014. He was using
all the so-called high-angular resolution techniques since it helped
his science to be made, namely study in details the inner structure of
the environments around stars, be it small mass (AGBs), more massive
(supergiant stars), or explosives (Novae). I will focus
here on his work on massive stars.
\end{abstract}
\maketitle
\section{Introduction}

The first published paper of Olivier Chesneau was on P Cygni
\citep{2000A&AS..144..523C}, the prototype Luminous Blue Variable star
(LBV), observed with one of the first astronomical adaptive optics
system, the \emph{Banc d'Optique Adaptative} at the \emph{Observatoire
  de Haute Provence} 1.52\,m telescope. He could resolve the
emission-line shell around this star and infer its mass loss. He later
started his PhD thesis on Wolf-Rayet stars because ``Wolf-Rayet''
sounded nice to his ears. He was a very nice person with a
never-ending curiosity, and his enthusiasm in peering into the physics
of stars marked all his collaborators.

\section{Why are massive stars interesting?}

We know today that some massive stars do explode as supernovae. As
such, they contribute to the kinematic feedback, compressing and
concentrating the interstellar medium (ISM) in some parts of the
Galaxy, and posing the initial conditions for future generations of
stars. Are all the massive stars progenitors of some supernovae? This
is being investigated nowadays and was one of the questions
O.~Chesneau wanted to tackle in the future.

\begin{figure}[htbp]
\centering
\frame{\includegraphics[width=10cm]{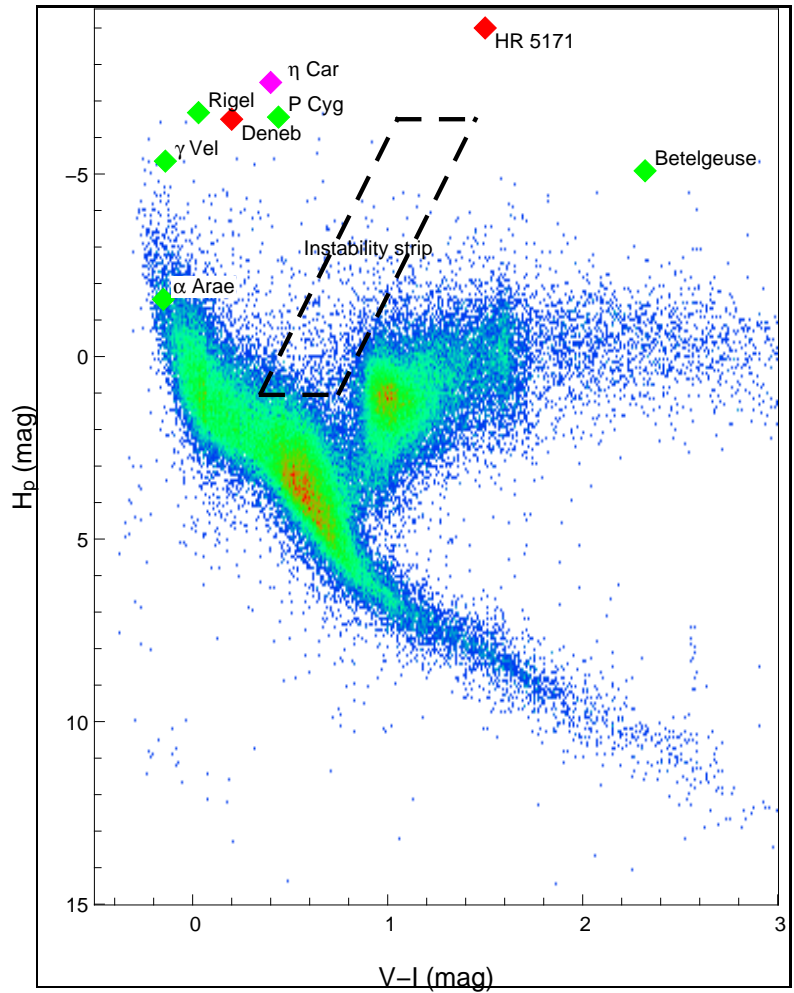}}
\caption{A Hertzsprung-Russel diagram generated from the Hipparcos
  catalog \citep{1997A&A...323L..49P}, in order to make it look like
  the one published on the ESA web-pages, showing the diagonal
  main-sequence of stars, and the upper-right branch of cool giant
  stars. The massive stars which were observed and published by
  O.~Chesneau are marked with Hipparcos-only colors and magnitudes
  (green), or with values from the literature (red and magenta). The
  instability strip, where the stars are pulsating or supposedly
  extremely short-lived, is marked in dashed line.}
\label{HR_Olivier}
\end{figure}

Massive stars form a very heterogeneous zoo, that astronomers have
arbitrarily classified as a function of temperature and
luminosity. They range from the low-temperature and high-luminosity
red supergiant stars (RSG), through yellow supergiant (YSG) and
hypergiant (YHG) stars, up to the high-temperature OB stars, be it on
the main sequence or more peculiar Wolf-Rayet stars (WR), Luminous
Blue Variable stars (LBV) or B[e] supergiant
stars. Figure~\ref{HR_Olivier} shows the massive stars that
O.~Chesneau focused on. They range from the cool supergiant star
Betelgeuse up to the hot Wolf-Rayet star Gamma Vel, all of them far up
in the Hertzsprung-Russel (HR) diagram.

Even during their short life (a few million years), these fat stars
have a tremendous mass loss, influencing their local environment and
sometimes triggering star formation in their vicinity. Detecting the
signatures of mass loss in massive stars formed the main scientific
playground of O.~Chesneau. It can manifest itself in the form of a
stellar wind and/or dust clumps that form in the nearest vicinity of
the star.

These stars also have very high luminosity in the range
$10^5-10^7$\,\Lsun which are the first source of ionization in our
Galaxy, revealing the interstellar clouds in fine and colorful laces
in star-forming regions. The high luminosity of massive stars helps in
forming dense and strong winds, but it may sound like an apparent
burden to form dust, which would be prevented from condensing by the
intense ultraviolet radiation of the central star. Aspherical
environments can help provide an explanation to dust formation in the
vicinity of such monsters. O.~Chesneau was fond of disks (as they are
relatively easy to detect with interferometry), and he was always
happy when he was discovering one around a star, but massive stars
have heavily challenged his -- and ours -- conceptions on their
close-by environments as asymmetries happen to be basically
everywhere.

I will present here only three prominent of O.~Chesneau's papers,
on which I had the chance to interact with him, be it a long time
after he published it (Eta Car) or for which I actually worked with
him.

\section[Eta Car]{Picturing the behemoth}

One of these challenging stars is Eta Carinae.

\begin{figure}[htbp]
\centering
\frame{\includegraphics[width=6cm]{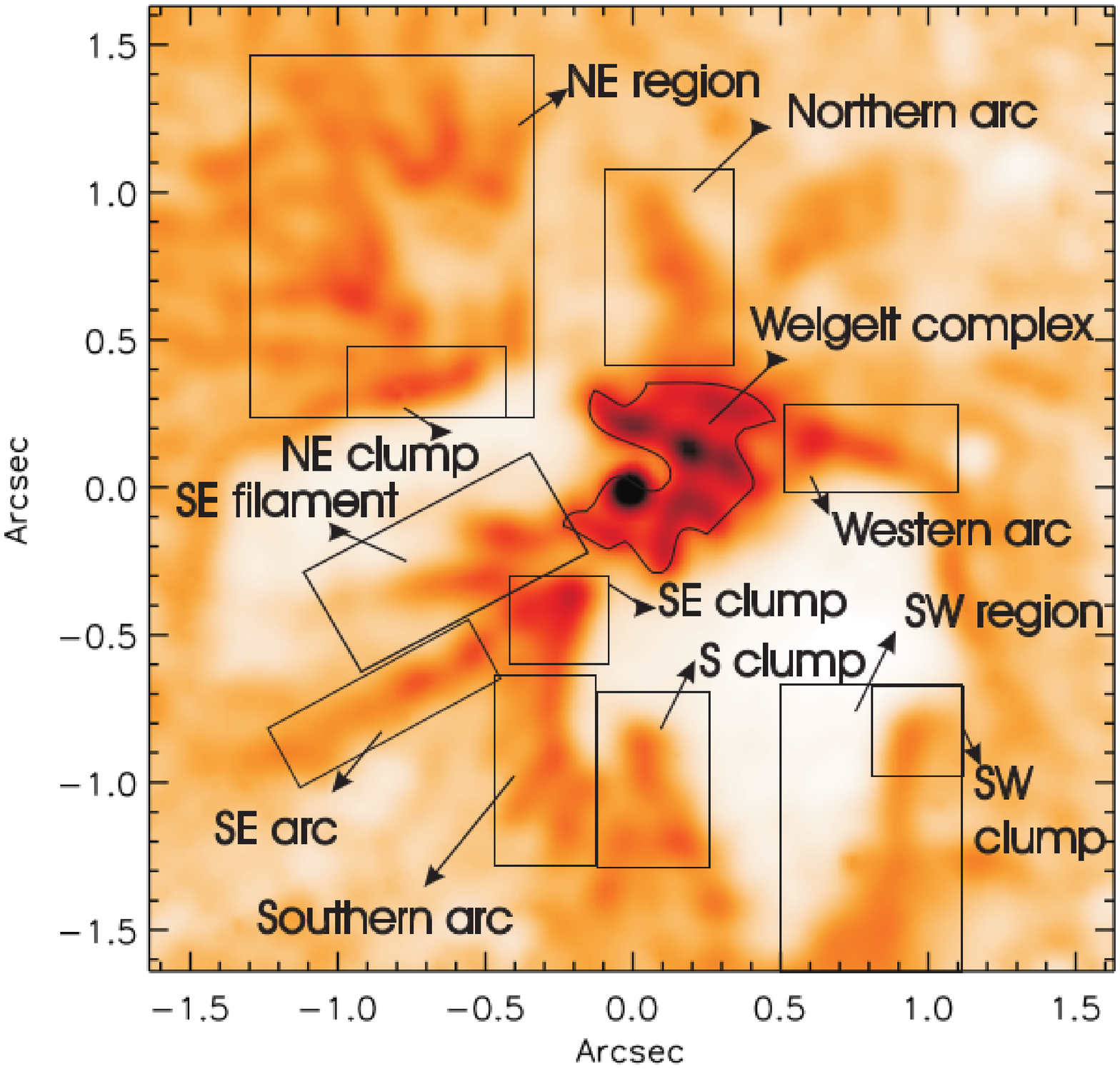}}
\frame{\includegraphics[width=6cm]{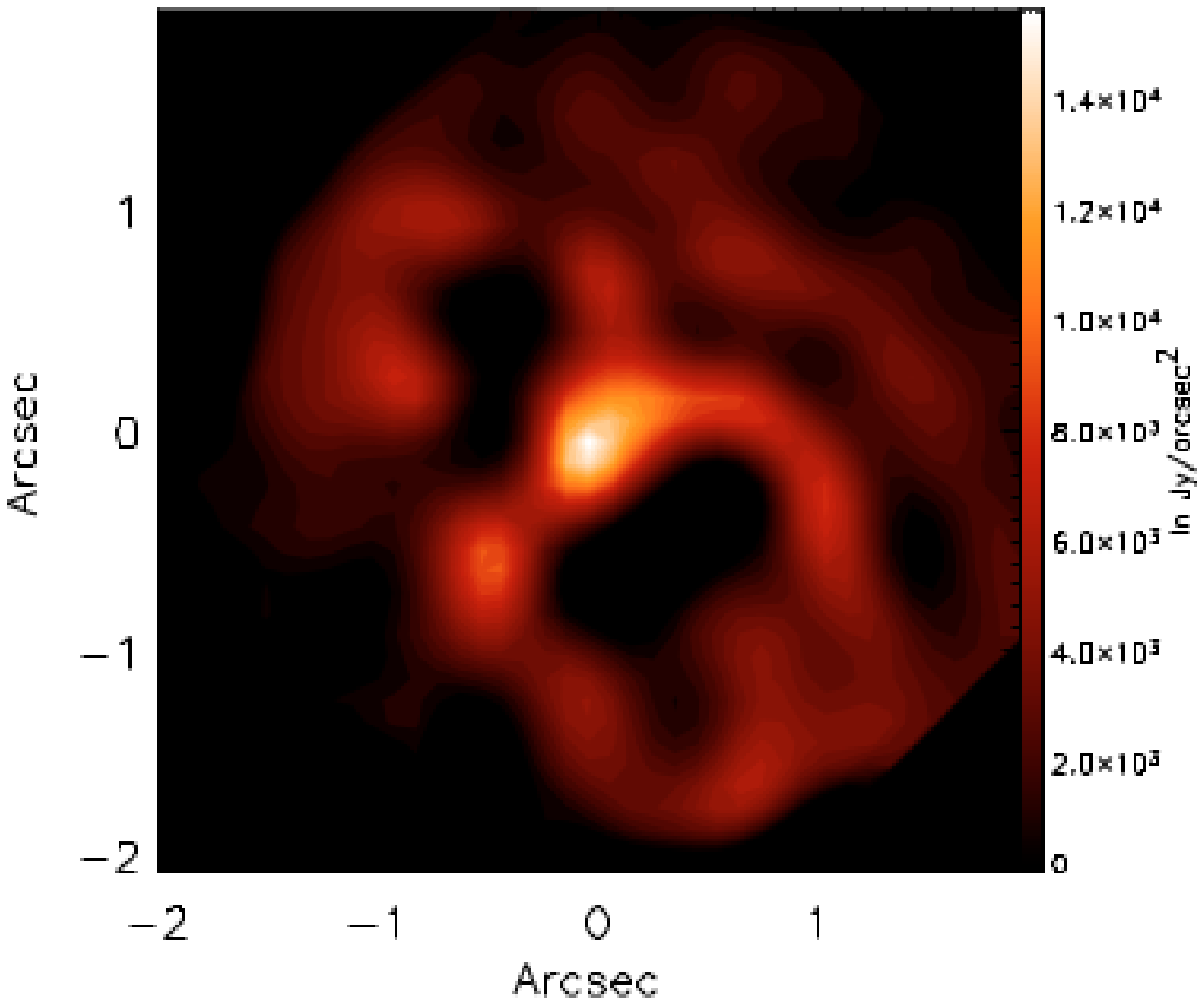}}
\caption{The ``Butterfly'' nebula within the Eta Car nebula from
  \cite{2005A&A...435.1043C}. {\bf Left:} In the near-infrared (with
  annotations). {\bf Right:} In mid-infrared.}
\label{EtaCar}
\end{figure}

I did not work with O.~Chesneau on this system, but I could get his
own impressions on it after the paper was published. He started
studying it through his implication in building the Mid-Infrared
(MIDI) instrument of the Very Large Telescope Interferometer (VLTI),
and he was trying to extract relevant information from the
observations he made. The multi-instrument observing campaign he
carried out on the system revealed a very complex inner structure,
where the dust form. The interpretation of these structures happened
to be very difficult for him. He finally could resolve the
dust-sublimation region around the central star and described the
complex morphology of the dust in its vicinity, showing for the first
time the so-called ``butterfly'' nebula, a dust-empty region in the
core of the larger scale Homunculus nebula (see Fig.~\ref{EtaCar}).

O.~Chesneau was later saying to the students in the laboratory, like
me, that he was directing his day-to-day research based on the lessons
learned from his work on Eta Car. Indeed, he was not well-prepared at
that time to make a complete study of the system. He could only be
descriptive in his article, but could not go further in the
interpretation due to the complexity of the object. Therefore, he
focused later to ``simpler'' objects like binaries and disks, for
which he had built very efficient observing methods and interpretation
tools. For every new study, when difficulties arouse in the
interpretation of the data, he was always saying in substance that
\emph{``this or that star is too complicated for me''}, but he always
was digging the subject to the very bottom and he was always finding
striking new properties on the stars he touched by means of his
favourite tools, i.e. adaptive optics with the Very Large Telescope
(VLT), and long-baseline interferometry with the VLTI array and the
CHARA array.

\section[HR5171a]{An unexpected binary monster}

O.~Chesneau was fond of all the objects with a high luminosity and a
high mass (but he was also fond of low-mass objects, see the previous
paper \citep{Lagadec2015}. Following his works on LBVs, he got
interested on the difference between LBVs and another type of
high-luminosity objects, namely yellow hypergiant stars. These stars
were thought to be some kind of more unstable (if it could be) LBV
stars, which are evolving very fast with time, skyrocketing to the
left (getting hotter) or to the right (getting cooler) in the HR
diagram.

O.~Chesneau looked into these extremely rare stars (about 10 are known
in our Galaxy) and found out that they were basically unstudied,
except from a few spectra and photometric data obtained a long time
ago on some of them. He therefore decided to get simple information on
them such as a diameter measurement, in order to clear the path of
future investigations. He made a telescope time proposal with the help
of S. Kanaan from Valparaiso to effectively measure the diameter of
two YHG stars V382 Car and HR5171a.

\begin{figure}[htbp]
\centering
\frame{\includegraphics[width=4cm]{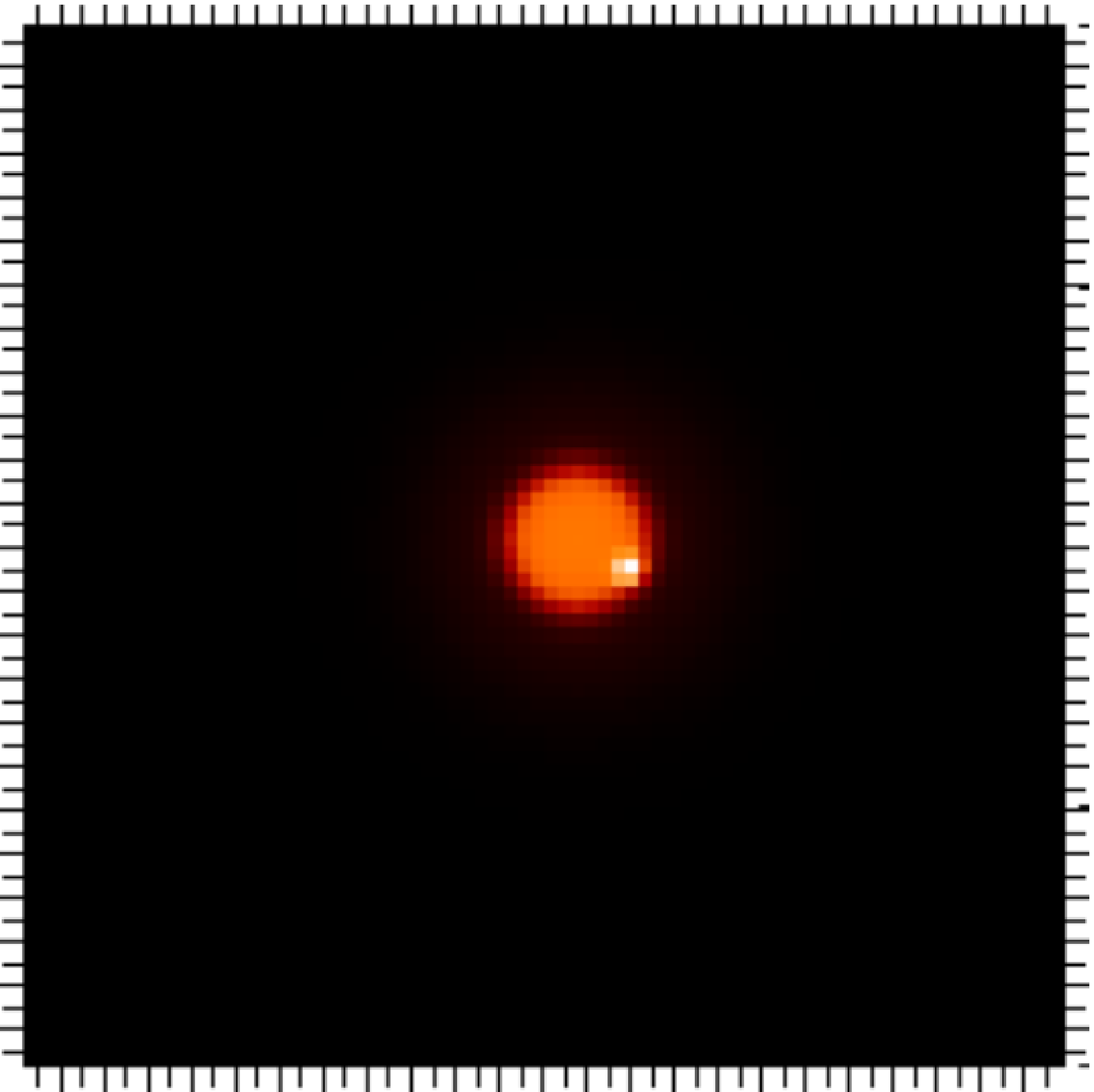}}
\frame{\includegraphics[width=8cm]{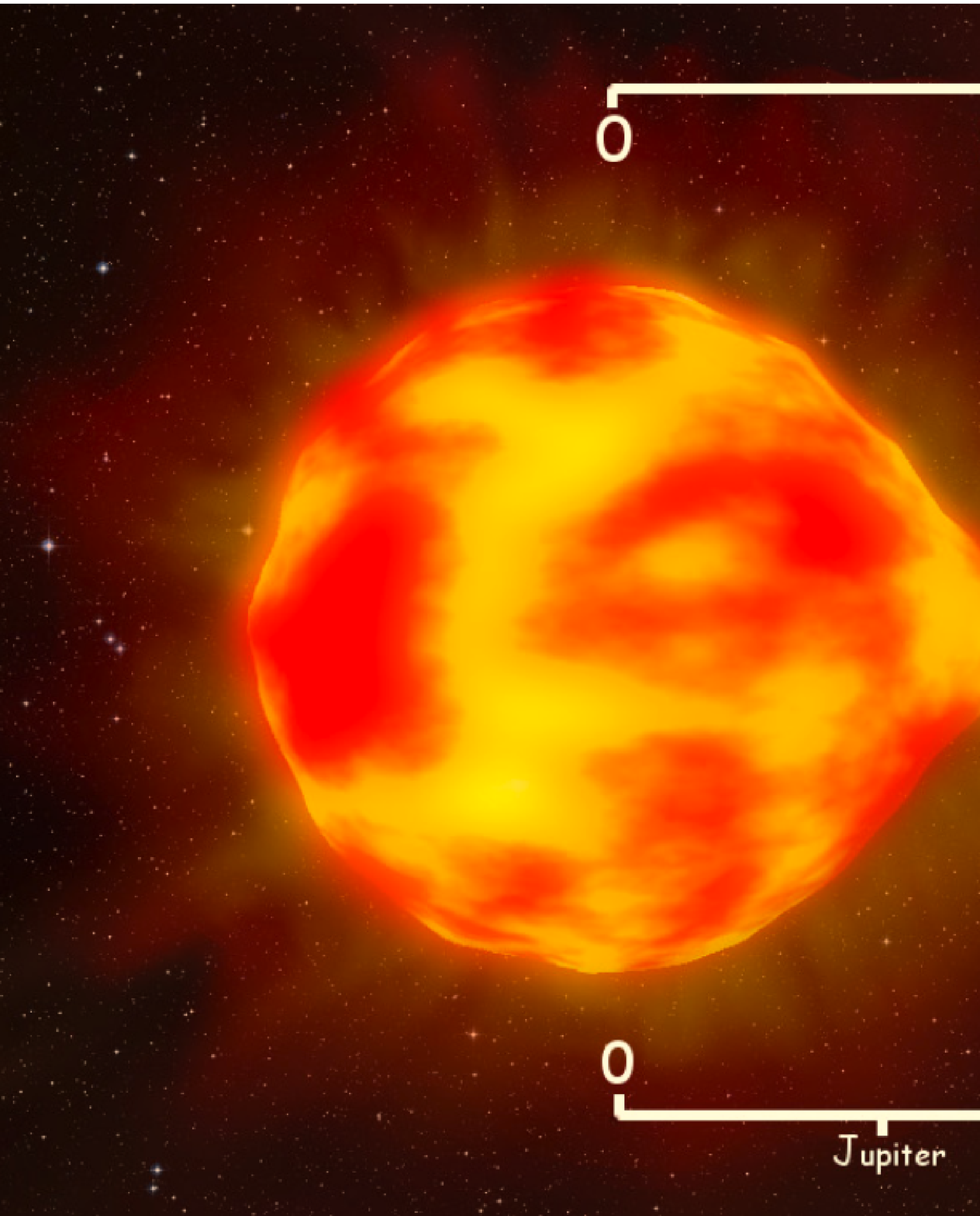}}
\caption{{\bf Left:} One of the first pictures made on HR5171a and
  circulated among co-authors. {\bf Right:} The final picture of the
  system based on the Roche-lobe overflow model of A. Meilland, as
  published \citep{Chesneau2014}.}
\label{HR5171a}
\end{figure}

The proposal was accepted and carried out by the VLTI. When looking at
the Astronomical Multi-Beam Recombiner (AMBER/VLTI) data, O.~Chesneau
could get a first interpretation on V382 Car, but the HR5171a data
resisted all attempts to use standard interpretation tools. This is
when he came to me with the data set he had, telling me in substance
\emph{``I tried many tools and ideas on this star but could not get a
  reliable diameter. I know you might have some tools which can be of
  help, could you please take a look to it?''}, a challenge which I
immediately accepted. When taking a look at the data on the star, I
noticed that it was not a plain single star, but probably made of an
unexpectedly large main star, in addition to an off-center bright
``blob'' or a companion star (see Fig.~\ref{HR5171a}). I remember the
excitement of O.~Chesneau to this unsuspected new fact and he
immediately set up a strategy to prove it was indeed a companion star,
by sending a message to his collaborators starting by the nice
sentence: \emph{``Florentin [...]  'sees' binaries everywhere... And
  he might be right!!!''}.

One peculiarity of this new found companion was the proximity to the
main star, nearly in contact with the main star
photosphere. O.~Chesneau therefore contacted photometric specialists to
check if there was a periodic variability, and indeed there was one!
This clarified substantially the picture of the system, which happened
to be formed of an extremely large star and a contact companion star,
in a so-called extremely rare common-envelope phase, where the
companion star is basically spiralling within the external layers of
the main star \citep{Chesneau2014}.  This kind of object can evolve
rapidly up until a merger event, like e.g. the V838 Mon event
\citep{Chesneau2014e}.

This is it on HR5171, because as O.~Chesneau wrote in an email to all
the co-authors in one of his typical enthusiastic messages,
\emph{``Explaining all the developments of the HR 5171 A project would
  be a (too) long (and wonderful!) story to tell.''}

\section[Gamma Vel]{Setting the distance to Gamma Vel}

I need to say a word on Gamma Vel before concluding this paper.

Gamma Vel is the closest WR+O star to Earth. As such, not only the
binary system but also the WR wind and shocked region between the WR
\& O winds could potentially be resolved by the VLTI. O.~Chesneau
set-up an observing program within the AMBER consortium to observe
this system, which was his ``pet'' star.

As early as the end of the first AMBER scientific night of
observation, in December 2004 (when Gamma Vel was first observed),
O.~Chesneau sent a warming congratulations message addressed to the
whole observing team: \emph{``Great!!!!!!!!!!!!! Congratulation to all of
you, I know how these observations can be stressful and exhausting. I
wish to the observer and the AMBER team other sucessfull nights and a
rich scientific harvest for AMBER. Good luck for the end of the run,
Olivier''}.

From the beginning, we were sure we had a wealth of information in the
Gamma Vel AMBER data, but we were puzzled by the interpretation
complexity of the spectro-interferometric measurements. Since
spectroscopy and interferometry were mixed for the first time in the
near infrared, we had to clear the path of interpreting such datasets.

When I moved to Nice in 2005, I started the real hard work on that
object (and others) with the wonderful enlightening of Olivier's
prolific ideas and help (with many visits to the then-separate
laboratories).

\begin{figure}[htbp]
\centering
\frame{\includegraphics[width=9cm]{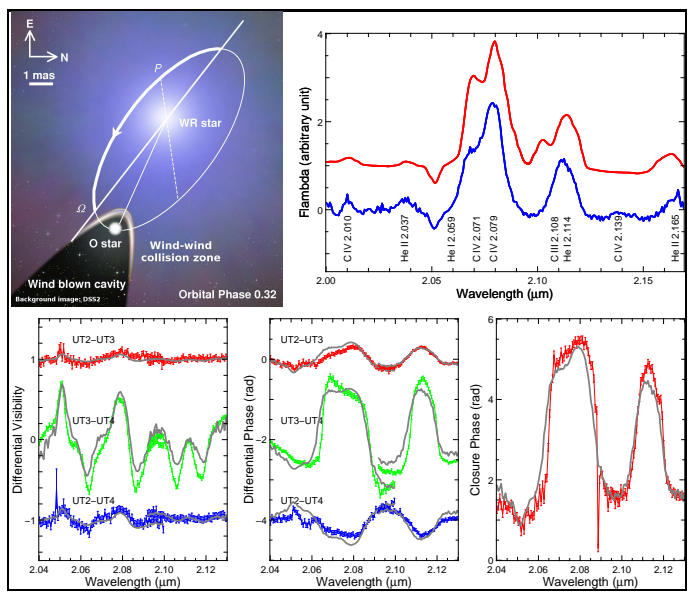}}
\caption{A summary of Gamma Vel observations (first published in the
  MPIfR Fachbeirat report in 2007). {\bf Top-Left:} Illustration of
  the system with the WR and O stars plus the wind-wind collision
  zone. {\bf Top-Right:} The observed spectrum (left) compared to the
  best-fit model (red). {\bf Bottom, from left to right:} Differential
  visibilities, differential phases and closure phases on Gamma Vel
  \citep{Millour2007}.}
\label{GammaVel}
\end{figure}

He was always coming up with bright ways of overcoming apparently
invulnerable boundaries in our work. I remember spending hours in
front of a whiteboard drawing sketches and writing equations together
to solve the interpretation problem we had on Gamma Vel data. The
pragmatism of Olivier finally took over and we ended up making a
simple geometrical model of the system, together with a thorough
investigation of other sources of information on the object, which
happened to learn us many things on the system: the distance we were
measuring on that target was at odds with the Hipparcos one, and we
could constrain the physical parameters of the WR wind.

The AMBER paper of Gamma Vel \citep{Millour2007} (see
Fig.~\ref{GammaVel}) was an excellent training on how to perform
research for the young student I was. O.~Chesneau's enthusiasm and
hard work were communicative and stimulating. I can write that the
exchanges with O.~Chesneau, starting from these ones on the Gamma Vel
work, were decisive in my life, as they convinced me that the field of
astrophysics and instrumentation research were what I wished to do
later.

\section{Concluding note: the impact of O.~Chesneau work on stellar physics.}

To illustrate the workforce of Olivier Chesneau, I will make here a
necessarily inaccurate and incomplete summary of his work. To give
numbers, he wrote or contributed to 97 peer-review articles, i.e. he
was contributing to the impressive number of 7 papers per year in
average (less than that at the beginning, closer to 10 papers per year
later). This represents about one out of four papers that were written
with VLTI data. At the date of today (2015), each of his articles
received about 25 citations in average.

Apart from rough numbers, he was heavily involved in the MIDI data
analysis software development, and as such he contributed to the two
highest impact papers of all history of interferometry, namely the
resolution of the dusty torus of an active galaxy nucleus
\citep{Jaffe2004}, and the discovery of radial segregation of dust in
protoplanetary disks \citep{vanBoekel2004b}.

On the hot stars side, his most prominent papers are on the behemoth
Eta Carinae \citep{2005A&A...435.1043C} and the fast rotator $\alpha$
Arae \citep{2005A&A...435..275C}, both setting a change in the vision
of these particular class of objects.

He was involved in national an international bodies to make better
science with the current facilities, among others ASHRA, EII, and
JMMC, and was also active in new projects like SPHERE and MATISSE.

Many people will remember O. Chesneau as a kind person, full of
enthusiasm. He touched many young researchers that will pursue his
work in the future.

\acknowledgements{The author would like to thank E. Lagadec,
  A. Meilland and L. Rolland for reading through this paper and making
  suggestions for improvements.}


\bibliographystyle{aa}
\bibliography{Chesneau}

\end{document}